\documentclass[]{spie}  

\usepackage{amsmath,amsfonts,amssymb}
\usepackage{graphicx}
\usepackage[colorlinks=true, allcolors=blue]{hyperref}
\PassOptionsToPackage{hyphens}{url}\usepackage{hyperref}
\usepackage{booktabs}
\usepackage{threeparttable, tablefootnote}
\title{SiAl Alloy Feedhorn Arrays: Material Properties, Feedhorn Design, and Astrophysical Applications}

\author[a,c]{Aamir M. Ali}
\author[b]{Thomas Essinger-Hileman}
\author[c]{Tobias Marriage}
\author[c]{John W. Appel}
\author[c]{Charles L. Bennett}
\author[b]{Matthew Berkeley}
\author[b]{Berhanu Bulcha}
\author[c]{Sumit Dahal}
\author[b]{Kevin L. Denis}
\author[b]{Karwan Rostem}
\author[b]{Kongpop U-Yen}
\author[b]{Edward J. Wollack}
\author[d]{Lingzhen Zeng}
\affil[a]{Dept. of Physics, University of California - Berkeley, Berkeley, CA 94720, USA}
\affil[b]{NASA Goddard Space Flight Center, Greenbelt, MD 20771, USA}
\affil[c]{Dept. of Physics and Astronomy, Johns Hopkins University, Baltimore, MD 21218, USA}
\affil[d]{Harvard-Smithsonian Center for Astrophysics, Boston, MA 02138}

\authorinfo{Further author information: (Send correspondence to Aamir M. Ali)\\Aamir M. Ali: E-mail: aamir.m.ali@berkeley.edu}

\begin{document} 
\maketitle

\begin{abstract}

We present here a study of the use of the SiAl alloy CE7 for the packaging of silicon devices at cryogenic temperatures. We report on the development of baseplates and feedhorn arrays for millimeter wave bolometric detectors for astrophysics. Existing interfaces to such detectors are typically made either of metals, which are easy to machine but mismatched to the thermal contraction profile of Si devices,  or of silicon, which avoids the mismatch but is difficult to directly machine. CE7 exhibits properties of both Si and Al, which makes it uniquely well suited for this application.

We measure CE7 to a) superconduct below a critical transition temperature, $T_c$, $\sim$ 1.2~K, b) have a thermal contraction profile much closer to Si than metals, which enables simple mating, and c) have a low thermal conductivity which can be improved by Au-plating. Our investigations also demonstrate that CE7 can be machined well enough to fabricate small structures, such as \#0-80 threaded holes, to tight tolerances ($\sim$ 25 $\mu$m) in contrast with pure silicon and similar substrates. We have fabricated CE7 baseplates being deployed in the 93 GHz polarimetric focal planes used in the Cosmology Large Angular Scale Surveyor (CLASS)\cite{Essinger-Hileman2014}. We also report on the development of smooth-walled feedhorn arrays made of CE7 that will be used in a focal plane of dichroic 150/220 GHz detectors for the CLASS High-Frequency camera.

\end{abstract}


\section{INTRODUCTION}
\label{sec:intro}  

Advancements in detector technologies have led to the fabrication of arrays of millimeter-wave monolithic detector arrays on wafers up to 200~mm in diameter. Two practical challenges to their use in astrophysical applications are 1) effective packaging in the cryogenic environments necessary to achieve stable, low-noise performance, and 2) an effective method of coupling light onto the detector. 

It is generally difficult to mate silicon devices to metallic substrates in cryogenic settings due to the significant mismatch in thermal contraction between silicon and typical metals. Mating large silicon wafers to metal is highly nontrivial and comes with a risk of damaging or destroying the silicon device. Using silicon itself as the packaging substrate is complicated due to the difficulty and cost of machining silicon and fabricating detailed features, such as threaded holes, feedhorn profiles in platelet stacks\cite{Britton2010,Simon2016}, etc. Some interfaces to Si are made with a Nickel-Iron metal alloy known as `Invar' (FeNi36) which shows minimal thermal contraction\cite{Nakamura1976}, but has the disadvantages of being highly ferromagnetic, which negatively impacts many detector technologies and their associated readout systems\cite{Ali2017, Vavagiakis2018}, and of being very soft for a metal and comparably more difficult to machine than aluminum or copper. Ceramics, such as Alumina, also have low thermal contraction but are also difficult to machine and highly insulating.

A proven technique for optical coupling is the use of feedhorns which mate to on-wafer antennas\cite{Leech2012, Simon2018}. Earlier work demonstrated feedhorns using a patented novel smooth-walled design that demonstrates the best non-corrugated feedhorn response to date with excellent beam symmetry, and low cross-polarization\cite{LingzhenZeng2010}. The use of a smooth wall and monotonic profile allows for efficient, cost-effective fabrication using relatively simple machining techniques, simpler than necessary for corrugated feeds\cite{Nibarger2012, Bersanelli1998, Torto2011} or silicon platelet arrays\cite{Simon2016,Nibarger2012} typically used in millimeter-wave applications. 

We present here our study of the Controlled Expansion 7 (CE7) SiAl alloy to address the general challenges of packaging silicon devices and our development of baseplates and smooth-walled feedhorn arrays made of CE7 for millimeter-wave detectors. CE7 is a proprietary Sandvik Osprey alloy consisting of 70\% Silicon and 30\% Aluminum, exhibiting a comparable coefficient of thermal expansion (CTE) to Si from room to cryogenic temperatures, and a superconducting transition at a critical temperature of 1.19 K, similar to the 1.18 K superconducting transition of Al. In-house Osprey fabrication capabilities enable the machining of pieces with detailed features to high tolerances.

This report will discuss several CE7 material properties, highlighting those that make it particularly advantageous for the cryogenic packaging of silicon devices. Details of experimental verification of these properties are presented. We  describe the machinability of the material as explored in the fabrication of baseplates and feedhorn arrays for millimeter-wave detectors, as well as the fabrication of a ring-loaded slot structure in the base of the smooth walled feedhorn which reduces return loss and facilitates the use of smooth-walled feedhorns over 100\% bandwidth.

\section{Material Properties}
\label{sec:material_prop}

\begin{table}[h!]
\centering
\begin{threeparttable}
\caption{ CE7 thermal, mechanical, and electrical properties, with comparisons to Si\cite{matweb_Si}, Al 6061\cite{ASM_Al6061}, Cu\cite{matweb_Cu}, and Alumina (Al$_2$O$_3$)\cite{matweb_alumina}, and measurements of CE7 mechanical properties were performed at room temperature by Sandvik Osprey\cite{sandvikCE}. Note that the similar values of the CE7 yield strength and ultimate tensile strength indicate the brittleness of the material, with minimal plastic deformation, a property inherited from silicon.}
\label{table:ce7_warm}
\begin{tabular}{l||l|l|l|l|l}
                                							& CE7       & Si    	& Al-6061 	& Cu    &	Al$_2$O$_3$\\ \hline
Ultimate tensile strength (MPa) 							& $\sim$100 & 7000  	& 310     	& 220   		& 260	\\
Yield strength (MPa)            							& 110       & 7000  	& 276     	& 33.3  		& 69	\\
Young's modulus (GPa)           							& 129.2     & 112.4 	& 68.0    	& 110   		& 370	\\
Rigidity modulus (GPa)          							& 51.6      & 43.9  	& 25.0    	& 46.0  		& 140	\\
Poisson's ratio                 							& 0.26      & 0.28  	& 0.36    	& 0.343 		& 0.22	\\
Density (g/cm$^3$)              							& 2.43      & 2.533 	& 2.699   	& 8.63 			& 3.90	\\
Superconducting $T_c$ (K)       							& 1.19      & N/A     	& 1.18    	& N/A 			& N/A	\\   
293~K $\Rightarrow$ 1~K Contraction ($\times 10^{4}$) \tnote{a} 	& $<$10  	& 2.16 		& 41.4 		& 32.6 			& 5.16	\\
Thermal Conductivity  300~mK (W / m $\cdot$ K)\tnote{b}          	& 0.02      & - \tnote{c}		& 0.85	  	& $\gtrsim$ 8.0 & 0.01	\\  
Resistivity 1.5 K ($\mu \Omega$ $\cdot$ m )\tnote{d}          			& 0.50      & -\tnote{c} 		&  1.5E-3   & 1.6E-4 		&  -\tnote{e}	\\  
\end{tabular}
\begin{tablenotes}\footnotesize
\item [a] Values for the linear thermal contraction calculated from 293~K to 1~K are mostly taken from White et al., 1993\cite{White1993}, while the value for Alumina (Al$_2$O$_3$) is taken from Wachtman et al., 1962\cite{WACHTMAN1962}.
\item[b] The thermal conductivity of Al 6061 and OFHC Copper (RRR=100) is taken from Woodcraft,\cite{Woodcraft2005} while the conductivity of Alumina is taken from the review by Simon et al. 1994\cite{Simon1994}.
\item[c]No value is given for the conductivity or resistivity of Si because these values vary considerably (particularly at low temperatures) depending on the composition of the silicon considered.
\item[d] Measurement of resistivity of Cu and Al 6061 taken from the review by Duthil\cite{Duthil2015} and assumes OFHC copper with RRR=100.
\item[e] No value of the resistivity is given for Alumina because it is extremely insulating at cryogenic temperatures ($2 \times 10^{12}$ $\Omega \cdot$m at room temperature, increasing monotonically with lower temperatures), and is better described by an electrical breakdown voltage (see, e.g., Simon et al. 1994\cite{Simon1994}).
\end{tablenotes}
\end{threeparttable}
\end{table}

As an alloy, CE7 maintains many of the material properties of silicon and aluminum. Its room-temperature properties are well documented and are summarized in Table~\ref{table:ce7_warm}. It is a dull gray, ceramic-like alloy, which is brittle, as indicated by the fact that the yield strength and ultimate tensile strength are essentially identical. It is lightweight and has similar Young's and rigidity moduli as Si. It is composed of interwoven continuous Si and Al phases, with a typical grain size $\sim 10$ $\mu$m.

To the best of our knowledge, the cryogenic properties of CE alloys relevant to the packaging of low-temperature silicon detectors have not been reported in the literature. Several cryogenic properties of CE7 were measured in an HPD Olympus\footnote{Model 104 Olympus Cryostat, High Precision Devices (HPD), 4601 Nautilus Ct S Boulder, CO 80301} adiabatic demagnetization refrigerator (ADR) cryostat at Johns Hopkins University. All temperature measurements were made using calibrated thermometers. Diode thermometers were used at temperatures above 4~K, while below 4~K either a calibrated Lake Shore\footnote{Lake shore Cryotronics, Westerville, OH 43082} germanium resistance temperature sensor (GRT) or calibrated Lake Shore Ruthenium Oxide (ROX) thermometers were used. Strain gauge and material resistances were read out using a Lake Shore LS370 resistance bridge.

\subsection{Superconductivity}

\begin{figure}
	\centering
    \includegraphics[width = .95
    \textwidth]{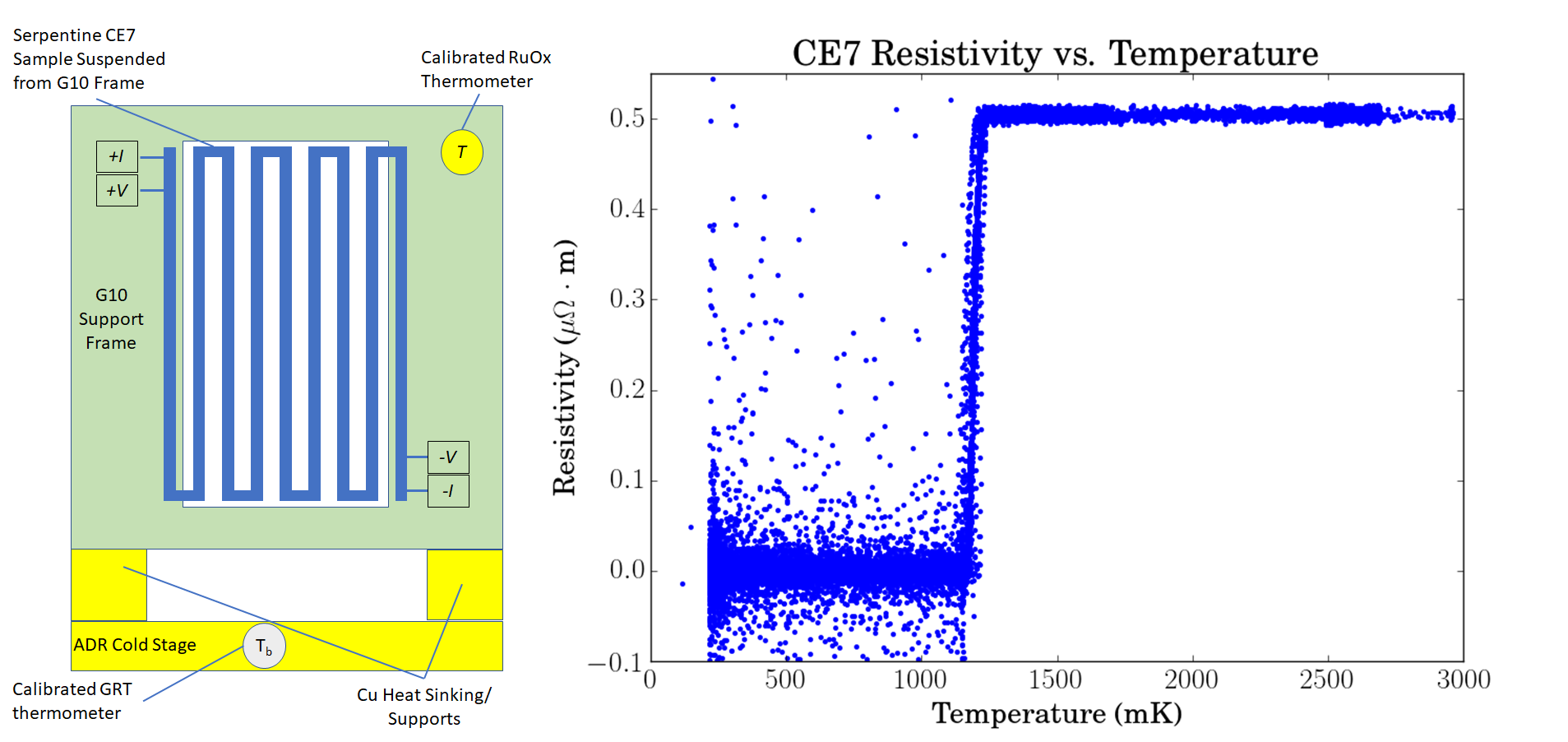}
	\caption{\textit{Left:} Schematic representation of CE7 resistivity test apparatus. Four leads were soldered to a serpentine meander of CE7 suspended from a G10 frame for electrical isolation. The resistivity of the sample was continuously measured from room temperature to 65~mK. The temperature of the sample was taken from a calibrated GRT thermometer mounted on the G10 frame, which was found to closely track the bath temperature ($T_b$) of the ADR cold stage. \textit{Right:} Measurements of the resistivity of CE7 at cryogenic temperatures. Note the superconducting transition with central critical temperature $T_c=1.19$~K, very close to the expected superconducting transition of Al around 1.18~K}
	\label{fig:CE7_tc}
\end{figure}

Cryogenic NbTi wire was soldered\footnote{During soldering, the CE7 sample was heated on a hot-plate to 150$^{\circ}$C, and soldered using a standard soldering iron at 300$^{\circ}$C using a high-purity 97-In 3-Ag solder.} in a 4-lead configuration across a `serpentine' sample of CE7 (cross sectional area 1.82 mm$^2$, length 1982 mm) with 26 meanders. The sample was mounted and heat sunk with 2 copper bars (on opposite sides of the sample) to the ADR cold stage. The temperature was PID servo controlled in 5~mK steps from 65~mK to 2.5~K to 65~mK 25 times. The resistance of the sample was recorded along with the temperature as measured by the GRT. The complete set of measurements of the 25 sweeps were averaged and binned at 5~mK intervals as shown in the right panel of Figure \ref{fig:CE7_tc}. 

The resistivity above $\sim$1.2~K is measured to be $0.50\pm.03$ $\mu \Omega \cdot$m. A super conducting transition from the full normal resistivity to 0 was observed with a width of roughly 90 mK, from 1.15~K to 1.24~K. Defined as the point where the resistivity is half of the normal resistivity, the critical temperature is measured to be $T_c = 1.190\pm.005$~K, remarkably close to the critical temperature of Al at 1.18~K\cite{Webb2015}.

\subsection{Thermal Contraction}

\begin{figure}[t]
\includegraphics[width = .95\textwidth]{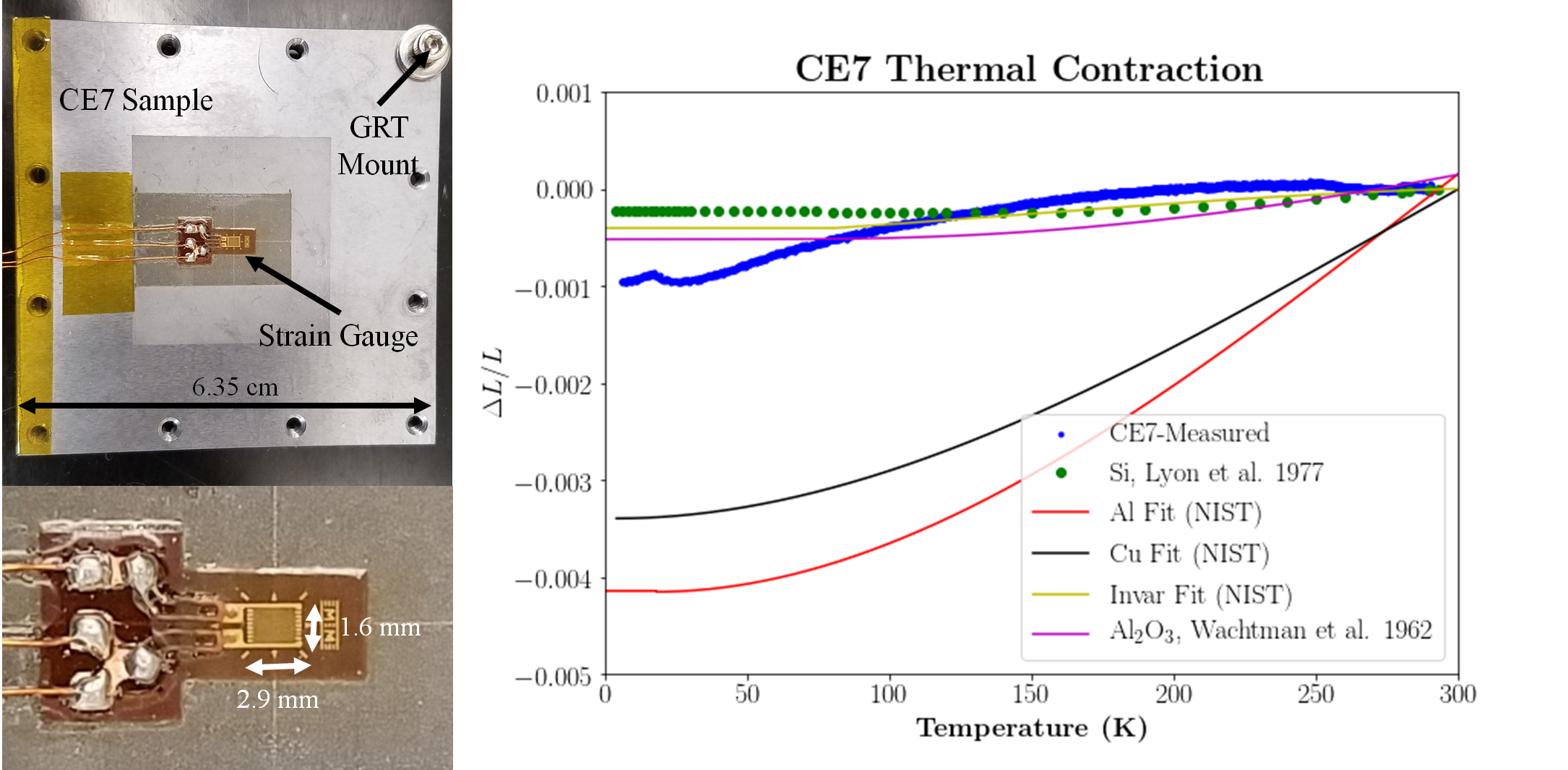}
\caption{ \textit{Top Left:} A calibrated strain gauge affixed to a sample of CE7, which was mounted directly to an ADR cold stage. The variation in CE7 color is the result of a roughening of the area to improve strain gauge adhesion. A GRT thermometer screwed into the CE7 sample measured the temperature, while the strain gauge measured the fractional contraction of the CE7 sample.\textit{Bottom Left:} Zoom in of strain gauge. \textit{Right:} Fractional contraction of CE7 as measured with a calibrated strain gauge (blue), with overlays of measurements of Si contraction as performed by Lyon et al. 1977\cite{Lyon1977} (green circles), and contraction curves of various substrates determined by the National Institute of Science and Technology\cite{Marquardt2000} such as Aluminum (red), OFHC Copper (black), and Invar (gold). A contraction curve of Alumina is taken from Wachtman et al., 1962\cite{WACHTMAN1962}} 
\label{fig:CE7_CTE}
\end{figure}

A Vishay WK-06-062AP-350\footnote{Vishay Precision Group, www.vishaypg.com} strain gauge was calibrated and epoxied to a 6.35 cm square sample of CE7 (thickness 6.25 mm) by PCM Measurements.\footnote{Proctor \& Chester Measurements Ltd., Warwickshire CV8 2UE UK} A measurement of the linear thermal contraction of CE7 (Figure \ref{fig:CE7_CTE}) was performed over a range of 300~K down to 4~K at Johns Hopkins. The fractional linear contraction was constrained to be less than 0.001. Figure \ref{fig:CE7_CTE} presents measurements of the fractional contraction overlaid with 1) measurements of Si contraction as performed by Lyon et al.\cite{Lyon1977} to cryogenic temperatures, and 2) curves of the contraction of Al, Cu, and Invar,  using polynomial fits to experimental data from the NIST cryogenic materials database\cite{Marquardt2000}. 

CE7 shows contraction more than a factor of 4 lower than Al\cite{Marquardt2000} and more than a factor of 3 lower than Cu\cite{Marquardt2000}, but approximately a factor 2 higher than the contraction of Alumina\cite{WACHTMAN1962}, about a factor of 3 higher than Invar\cite{Marquardt2000}, and a factor of 5 higher than the contraction of Silicon as measured by Lyon et al.\cite{Lyon1977} The discrepancy between CE7 and Silicon is due to the significant presence of Al in CE7, which contracts by $\Delta L/L \sim4 \times 10^{-3}$ over this temperature range. A simple weighted mean of the contraction of 70\% Si and 30\% Al gives a predicted contraction of $\sim.0012$, very close to the observed contraction of CE7. 

\subsection{Thermal Conductivity}

Below the superconducting transition of 1.19 K, CE7 is expected to be a relatively poor thermal conductor, with a conductance worse than silicon. The thermal conductivity was measured using a rectangular bar (approximately 10 cm $\times$ 1 cm $\times$ 1 cm) mounted on the bottom to the 100 mK ADR cold stage, and a heater of known resistance mounted on the top to supply thermal power. The thermal gradient was measured using 5 calibrated RuOx thermometers screwed into the CE7 distributed evenly along the bar (1.66 cm separation). The power applied to the sample varied between 0 and 14 $\mu$W, yielding data on the thermal conductivity of CE7 at temperatures below 500~mK. Given the low conductivity of CE7, an identical bar was prepared and plated with gold and analogously measured in order to investigate if gold plating improved the conductivity (see section \ref{sec:gold_plating} for details of plating procedure).

\begin{figure}
	\centering
    \includegraphics[width = .95\textwidth]{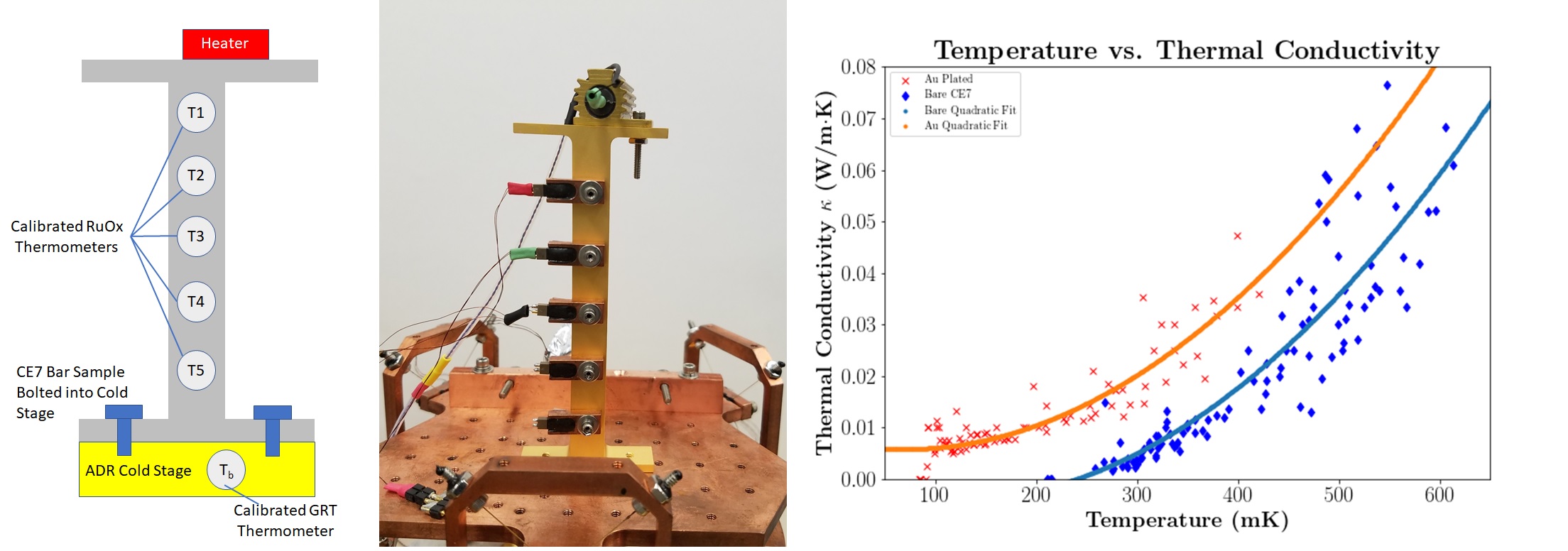}
	\caption{\textit{Left:} Schematic representation of CE7 thermal conductivity measurement. The bottom of a bar of CE7 was screwed to the cold stage of an ADR (thermal sink), while a heater was screwed to the top of the bar. Along the bar, 5 RuOx thermometers were screwed directly into the CE7 to measure a temperature gradient induced by the heat load. \textit{Center:} A photograph of an Au-plated CE7 bar thermal conductivity test setup, as described. Measurements of the thermal conductivity were performed over multiple cooldowns on both bare and Au-plated CE7 bars. \textit{Right:}  Measurements of the thermal conductivity $\kappa$ of bare CE7 (orange x) and Au-plated CE7 (blue diamond) at various temperatures, with quadratic fits (equations \ref{eq:kappa1} and \ref{eq:kappa2}) overlayed.}
    \label{fig:temp_vs_k}
\end{figure}

Below, 500~mK, the thermal conductivity of CE7 and Au-plated CE7 can be approximated by simple 2nd order polynomials:
\begin{align}
\kappa \left(\text{CE7 Bare}\right) &= .27 \, T^2 - \left(6.2\times 10^{-2}\right) T - 7.3 \times 10^{-4} \label{eq:kappa1}\\
\kappa \left(\text{CE7 Au Plate}\right) &= .27 \, T^2 - \left(3.7\times 10^{-2}\right) T + 7.1 \times 10^{-3} \label{eq:kappa2}
\end{align}
with the thermal conductivity, $\kappa$ given in units of \mbox{$W/ m \cdot K$}, and the temperature of the substrate $T$ given in degrees K. As seen in the right panel of Figure \ref{fig:temp_vs_k}, at temperatures below $\sim$ 200~mK, the polynomial fit predicts an unphysical negative conductivity. This is likely due to the breakdown of the assumption of a smoothly varying thermal conductivity across the length of the bar in our analysis, and the fit should be considered inaccurate below $\sim$ 250~mK.

Two common operating temperatures for millimeter wave detectors are 300~mK and 100~mK, corresponding to the usual base temperatures of  He adsorption refrigerators for the former and ADR/DR refrigerators for the latter. The measured thermal conductivity of bare CE7 at 300~mK is approximately 0.005$\pm$0.001  ($W/ m\cdot K$) while the bare CE7 was unable to cool to 100 m$K$ ($W/ m\cdot K$). This is likely due to what we measure to be a vanishing thermal conductivity near 220~mK. The equivalent thermal conductivity for Au-plated CE7 is 0.02$\pm$ 0.005 ($W/ m\cdot K$) at 300~mK and 0.006$\pm$0.002 ($W/ m\cdot K$) at 100~mK. The observed increase in conductance of the sample with gold electroplating is accounted for given the thickness deposited if the gold had a RRR$\sim$5\cite{Rosenberg1955}, a plausible value in the presence of impurities in the plated gold.


\section{CE7 Detector Interface Plates and Feedhorns}
\label{sec:discussion}

\begin{figure}[t!]
	\centering
    \includegraphics[width = .95\textwidth]{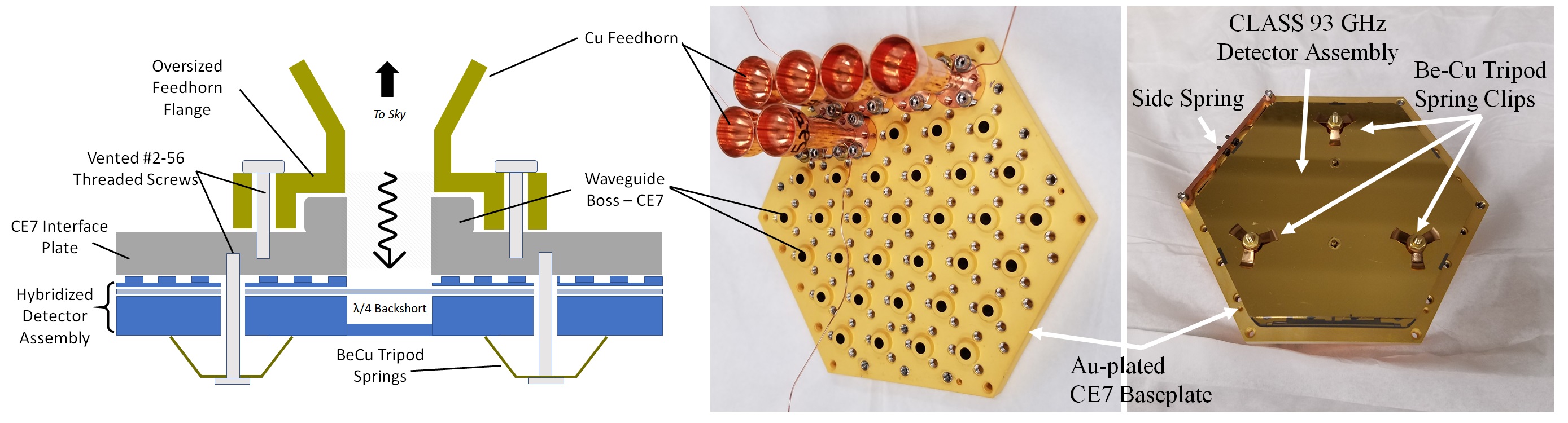}
	\caption{CLASS 93 GHz Au-plated CE7 waveguide interface plate, mating CLASS 93 GHz detector assembly with Cu feedhorns \textit{Left:} Schematic representation of interface between detector assembly, CE7 interface plate, and Cu feedhorn. The interface plate consists of an array of cylindrical boss extrusions housing the waveguides to interface with the Cu feedhorns. A corresponding feature at the base of the Cu feedhorns is oversized so as to `grab' the cylindrical boss with the cryogenic contraction of the Cu. The feedhorns are screwed into threaded \#2-56 holes machined in the CE7 plate with inserted helicoils. The detector assembly is vertically constrained simply with the use of three Be-Cu spring clips screwed into threaded \#2-56 holes in the CE7 plate. \textit{Center} Photograph of CE7 interface plate, partially populated with Cu feedhorns. \textit{Right} A CLASS 93 GHz detector assembly is mounted onto the CE7 interface plate. Waveguide alignment ($x$-$y$ plane) is accomplished by means of 2 alignment pins situated in the baseplate which are designed to press against a flat edge and corner groove in holes in the detector wafer. A side mounted Be-Cu spring applies between .25-0.5 N of force to press the alignment pins against the hole features. This constrains the rotation and translation of the wafer ensuring waveguide to antenna-pair alignment}
    \label{fig:W_band_mounting}
\end{figure}

The results presented in section \ref{sec:material_prop} have implications for the suitability of CE7 for the cryogenic packaging of Si devices, particularly for astrophysical waveguide interfaces and feedhorn arrays. Based on these results, CE7 is employed as the packaging material for detectors in 3 of the 4 telescopes in the Cosmology Large Angular Scale Surveyor (CLASS) Cosmic Microwave Background (CMB) telescope array \cite{Ali2017, Dahal2018}. For the two CLASS 93 GHz telescopes (the first of which is deployed), interface plates with circular waveguides are made of CE7 and couple/align the detector array wafers to copper feedhorns and provide the base for the overall structure of the detector modules (Figure \ref{fig:W_band_mounting}). For the CLASS dichroic 150/220 GHz telescope, the baseplate will be a feedhorn array machined out of a monolithic block of CE7. The ensuing discussion will address the suitability of CE7 for Si packaging, leveraging both the measurements made above as well as the practical experience gained by its use in the CLASS telescopes. Relevant comparisons are made to other substrates.

\subsection{Superconductivity and Shielding}
\label{sec:shielding}
It is worth mentioning that since CE7 superconducts below 1.2 K, it will behave as a magnetic shield for the detector/readout systems. The transition edge sensor (TES) bolometers and superconducting quantum interference device (SQUID) amplifiers in the detector assemblies are both susceptible to magnetic fields, and superconductors have been used in similar architectures to provide magnetic shielding for analogous astrophysical applications\cite{Hollister2008}. Since detector assemblies cannot be enclosed by shielding in the direction of light from observations, an element of shielding on this plane of the detector is likely to improve the magnetic environment stability, although this effect needs to be studied. By comparison, Si, Cu and Alumina do not superconduct, which preclude such a shielding advantage, while the presence of Invar, which is ferromagnetic, may intensify the magnitude of the magnetic field and may interact with ambient fields in an undesirable way.

\subsection{Thermal Conductivity and Gold Plating}
\label{sec:gold_plating}
Due to the low thermal conductivity of CE7 at cryogenic temperatures, it was found necessary to gold plate parts to ensure that they thermalized properly. The plating process that we found successful was 7 $\mu$m electroless nickel followed by 3.8 $\mu$m soft gold.\footnote{Per ASTM B 488 Type III Grade A.} We investigated gold plating with non-magnetic materials, particularly a Ag underplating. We found this plating procedure to be less reliable on average than the Ni underplate, occasionally resulting in surface blistering, resulting in the selection of the comparably more reliable Ni underplate. Transition Edge Sensor (TES) detectors were characterized both with the Ni underplate and without to see if the magnetic material shifted the detector properties (most notably, the $T_c$ of the TES) but we found no evidence of any shift at the 3 mK level. This indicates that the presence of the thin Ni underplate does not disrupt the magnetic field in the vicinity of the devices significantly.  

Cu is a more desirable choice of substrate purely from the perspective of thermal conductivity. Al is also thermally conductive, but much less than Cu from the limited conductivity data available at very low temperatures\cite{Woodcraft2005}. Both metals have a large thermal contraction mismatch with Si, as has been mentioned. In practice, Si, Invar, and Alumina, all have low thermal conductivities and are often plated (or otherwise) if thermal conductance is a driving consideration.

\subsection{Thermal Contraction}
Although the contraction of CE7 is greater than pure Silicon, it is a much closer match than metallic substrates, and is acceptable given the gain in fabrication ease vs. silicon (section \ref{sec:fab}). In practice, 93 GHz CLASS detectors have been mounted simply to CE7 baseplates (Figure \ref{fig:W_band_mounting}), and warmed/cooled over multiple cycles ($>$10) with no evidence of misalignment or other failure. Thermal contraction needed to be accounted for primarily at interfaces where the CE7 would mate to metals, rather than to the detector silicon itself, although for higher precision alignments, the differential contraction may need to be accounted for.

The thermal contraction of CE7 is far less than comparable metals, and is slightly higher than Alumina, Invar, or Si. We have found CE7 to be a sufficiently close match at the tolerances of interest to not drive a choice to the latter three substrates: Alumina and Si were down-selected due to machinability (Section \ref{sec:fab}) and cost concerns, while Invar was rejected largely due to magnetic concerns (Section \ref{sec:shielding}).   

\begin{figure}
	\centering
    \includegraphics[width = .95\textwidth]{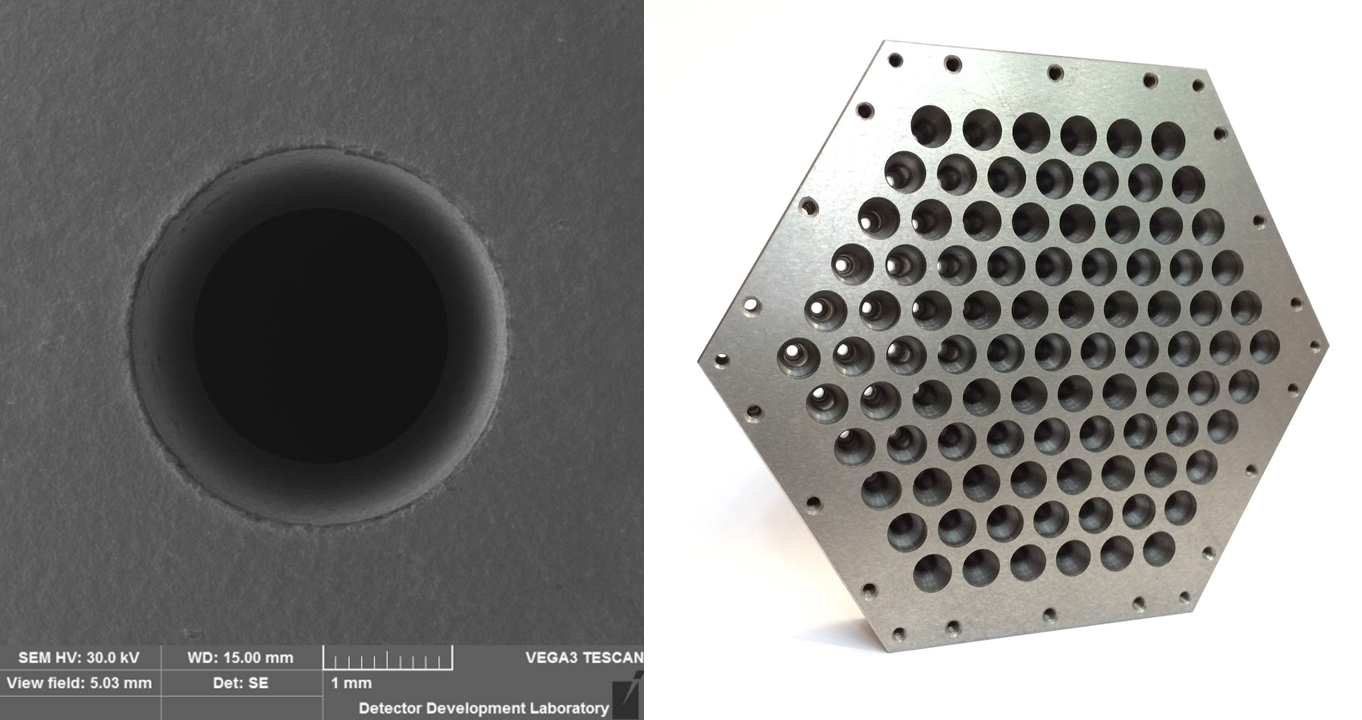}
	\caption{\textit{Left:} Scanning Electron Micrograph of 93 GHz CLASS detector baseplate waveguide fabricated out of CE7. \textit{Right:} Photograph of a 91-element CE7 feedhorn array. This array was machined with conservative spaces between the tops of feedhorns. Parts with overlapping feedhorns have been produced.}
    \label{fig:ring_load}
\end{figure}

\subsection{Fabrication}
\label{sec:fab}
CE7 is machinable in a manner easier than ceramics or silicon, but not as easy as typical metals. The characteristic Si grain size of the alloy is of order 10 $\mu$m.\footnote{From Robert Ross of Sandvik Osprey. Personal communication, July 4th, 2017} Structures approaching this size are typically difficult to machine because fractures occur in the material at grain boundaries; however, parts have been produced with surface finish better than 0.4 $\mu$m and flatness better than 5 $\mu$m. Sandvik Osprey has demonstrated the ability to machine threaded holes as small as \#0-80 (with feature tolerances of the threads of order 0.01 mm). The waveguides fabricated in the CLASS 93 GHz interface plates held diameter tolerances of $\pm$ 10 $\mu$m, and the waveguide boss extrusions were fabricated to tolerances of $\pm$ 25 $\mu$m. Tolerances of $\sim$ 25 $\mu$m is hard, but doable, while tolerances $\sim$ 10 $\mu$m is near the edge of capability, and can be achieved with specialized machining techniques, like using precision reamers or plunge electrical discharge machining.

Features approaching a few times the grain size were most successfully machined using a plunge electrical discharge machining process, which avoids placing undo stress on the material during machining. As an example, a ring-loaded impedance-matching section was successfully machined into a CE7 waveguide section at the base of a feedhorn, as shown in Fig.~\ref{fig:ring_load}. The ring-loaded structure contained a raised boss 93~$\mu$m tall and 208~$\mu$m wide. This feature was successfully machined, but features much smaller would be very challenging. 

\begin{figure}[t!]
	\centering
    \begin{tabular}{cc}
    	\includegraphics[height = .26\textwidth]{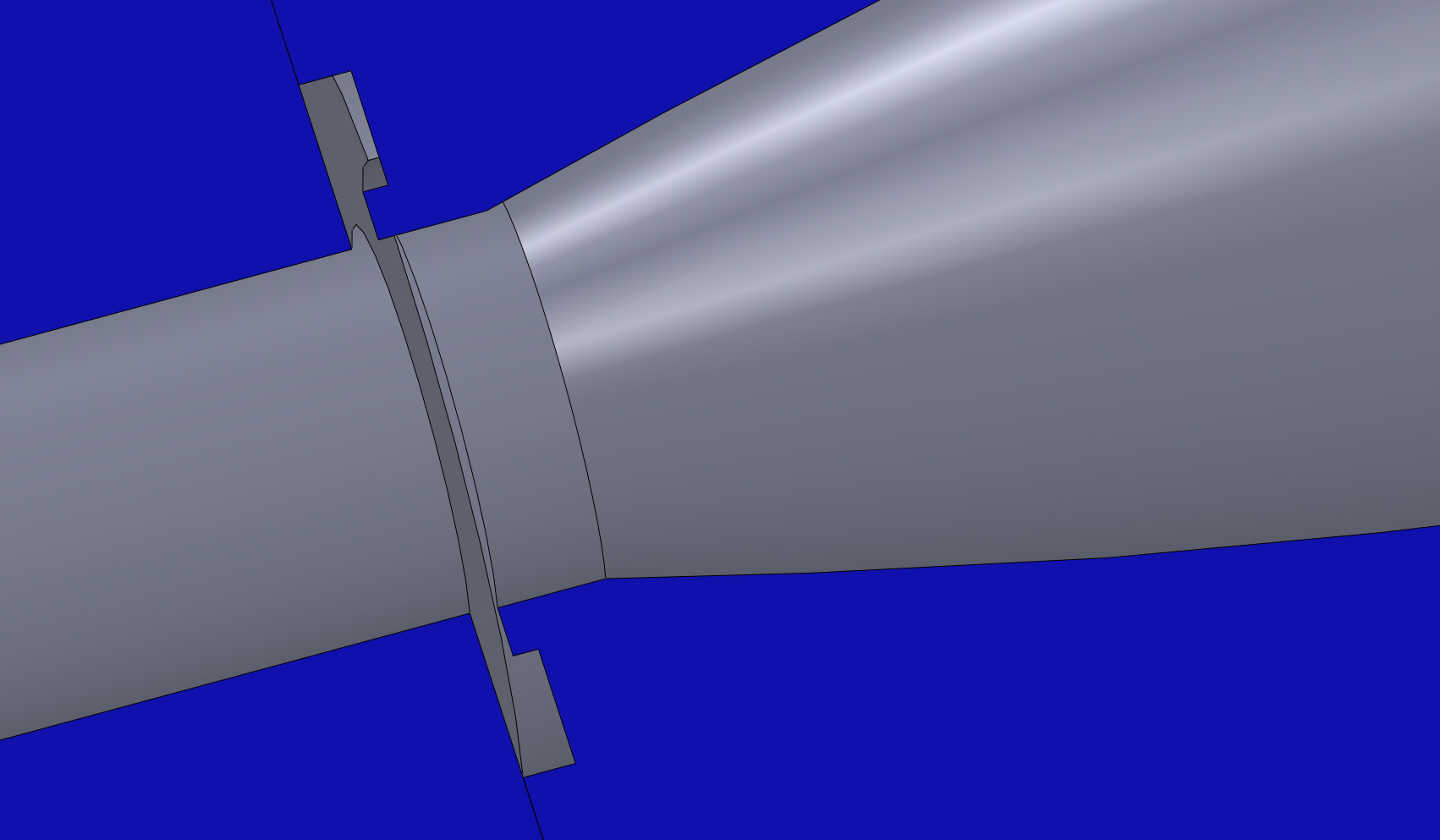} &
        \includegraphics[height = .26\textwidth]{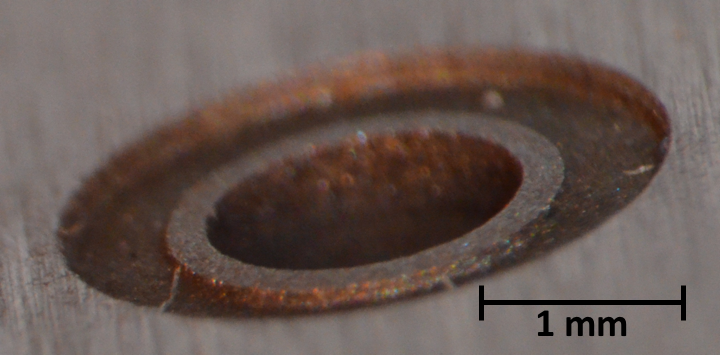}
    \end{tabular}
	\caption{\textit{Left:} Model of a ring-loaded impedance-matching structure at the base of a feedhorn. \textit{Right:} Photograph of the structure realized in CE7 through a plunge electric discharge machining (EDM) process, showing the ability to manufacture fine features in CE7. }
    \label{fig:ce7_hfarray}
\end{figure}

\section{Conclusion}

We have reported on the cryogenic material properties of CE7 and commented on the suitability of CE7 for mating to silicon devices in astrophysical settings. In particular, we have characterized the low temperature resistivity and a super conducting transition at 1.19 K, measured the thermal contraction to be $\Delta L / L \simeq$ 0.005 from room temperature to 4~K, and measured a bare CE7 and Au-coated CE7 to have a thermal conductivities of 0.005$\pm$0.001 and 0.02$\pm$ 0.005 ($W/ m\cdot K$) respectively at 300~mK, with a vanishing thermal conductivity for the bare CE7 below $\sim$ 220~mK. In Section \ref{sec:discussion} we described the suitability of CE7 for packaging Si in cryogenic contexts. We also fabricated waveguide array interface plates and feedhorn arrays out of CE7 for use in the CLASS CMB telescopes and described the mating procedure, practically demonstrating CE7 to be a mature and attractive choice for astrophysical instruments.

\acknowledgments
The authors graciously acknowledge a private gift from Matthew Polk (Johns Hopkins Physics and Astronomy B.S. alum, 1971) which funded much of this investigation, including Aamir Ali for a portion of his graduate studies. We acknowledge the National Science Foundation Division of Astronomical Sciences for their support of CLASS under Grant Numbers 0959349, 1429236, 1636634, and 1654494. CLASS uses detector technology developed under several previous and ongoing NASA grants. Detector development work at JHU was funded by NASA grant number NNX14AB76A. We are also grateful to NASA for their support of civil servants engaged in state-of-the-art detector technologies. T. Essinger-Hileman was supported by an NSF Astronomy and Astro-
physics Postdoctoral Fellowship.

 \bibliography{report} 
 \bibliographystyle{spiebib} 

\end{document}